\documentclass[twocolumn,preprintnumbers,amsmath,amssymb]{revtex4}

\usepackage[dvipdfmx]{graphicx}
\usepackage{bm}
\usepackage{color}
\usepackage{amsmath}
\usepackage{amsfonts}
\usepackage{amssymb}
\usepackage{ulem}
\usepackage{array}
\usepackage{setspace}

\begin{document}

\preprint{APS/123-QED}

\title{Magnetic phase diagram in three-dimensional triangular-lattice antiferromagnet Sr$_3$CoTa$_2$O$_9$ with small easy-axis anisotropy}

\author{Iori Nishizawa$^1$}
\author{Nobuyuki Kurita$^1$}
\email{kurita.n.aa@m.titech.ac.jp}
\author{Hidekazu Tanaka$^{1,2}$}
\email{tanaka.h.ag@m.titech.ac.jp}
\author{Takayuki Goto$^3$}

\affiliation{$^1$Department of Physics, Tokyo Institute of Technology, Meguro-ku, Tokyo 152-8551, Japan\\
$^2$Innovator and Inventor Development Platform, Tokyo Institute of Technology, Midori-ku, Yokohama 226-8502, Japan\\
$^3$Physics Division, Sophia University, Chiyoda-ku, Tokyo 102-8544, Japan}

\date{\today}

\begin{abstract}
We report the results of low-temperature magnetization and specific heat measurements of Sr$_3$CoTa$_2$O$_9$ powder, in which Co$^{2+}$ ions with effective spin--1/2 form a uniform triangular lattice in the $ab$ plane. It was found that Sr$_3$CoTa$_2$O$_9$ undergoes successive antiferromagnetic transitions at $T_{\rm N1}\,{=}\,0.97~{\rm K}$ and $T_{\rm N2}\,{=}\,0.79~{\rm K}$ at zero magnetic field. As the magnetic field increases, both $T_{\rm N1}$ and $T_{\rm N2}$ decrease monotonically. The obtained magnetic field vs temperature phase diagram together with a sharp magnetization anomaly at a saturation field of ${\mu_0}H_{\rm s}\,{=}\,2.3$\,T indicates that Sr$_3$CoTa$_2$O$_9$ is described as a spin--1/2 three-dimensional triangular-lattice antiferromagnet with a weak easy-axis anisotropy. We discuss the characteristics of the magnetic phase diagram, which approximates the phase diagram for the magnetic field perpendicular to the $c$ axis.
\end{abstract}


\maketitle


\section{Introduction}\label{Intro}
Triangular-lattice antiferromagnets (TLAFs) are one of the most widely investigated spin models. TLAFs show a variety of phase transitions in magnetic fields depending on magnetic anisotropy, spatial anisotropy, and interlayer exchange interaction~\cite{Collins,Balents,Starykh2}. Two-dimensional (2D) classical Heisenberg TLAFs with the easy-axis anisotropy display a magnetization plateau at one-third of the saturation magnetization when a magnetic field is applied along the easy-axis~\cite{Miyashita}. The classical magnetization plateau vanishes in the absence of anisotropy at $T\,{=}\,0$. For quantum spin, such as spin--1/2, 2D Heisenberg TLAFs exhibit a 1/3-magnetization-plateau without the help of the easy-axis anisotropy, where the up-up-down (uud) spin state is stabilized by quantum fluctuation in a finite magnetic field range~\cite{Nishimori,Chubokov,Nikuni,Honecker,Alicea,Farnell,Sakai,Hotta,Yamamoto1,Sellmann,Coletta}. At finite temperatures, not only quantum fluctuation but also thermal fluctuation stabilizes the uud state so that the field range of the uud state increases with increasing temperature~\cite{Kawamura,Lee_theory,Seabra,Gvozdikova}. 

The interlayer exchange interaction is crucial for the emergence of quantum phases, including the 1/3-magnetization plateau~\cite{Gekht}. Real TLAF compounds have finite interlayer exchange interactions. When the interlayer exchange interaction is ferromagnetic, the macroscopic quantum effect is robust as observed in CsCuCl$_3$~\cite{Nojiri_CsCuCl3,Sera}, in which the ferromagnetic interlayer exchange interaction is approximately six times larger than the antiferromagnetic intralayer exchange interaction~\cite{Nikuni,Tazuke,Tanaka_JPSJ1992}. When the interlayer exchange interaction is antiferromagnetic and weak, a cascade of quantum phase transitions occurs in magnetic fields as observed in Cs$_2$CuBr$_4$~\cite{Ono1,Ono2,Fortune} and Ba$_3$CoSb$_2$O$_9$~\cite{Shirata,Zhou,Susuki,Quirion,Koutroulakis,Yamamoto2,Okada}. As the ratio of the interlayer exchange interaction to the intralayer exchange interaction increases, macroscopic quantum effects are wiped out, and the system behaves as a classical TLAF~\cite{Li}. However, the magnetic field vs temperature phase diagram has not been sufficiently understood even in the case of a classical TLAF, although a qualitative phase diagram was discussed phenomenologically~\cite{Plumer_PRL1988,Plumer_PRB1989}. Thus, it is worthwhile investigating phase diagrams in TLAFs with various interlayer exchange interactions. 

\begin{figure}[t]
	\includegraphics[width=7.0cm]{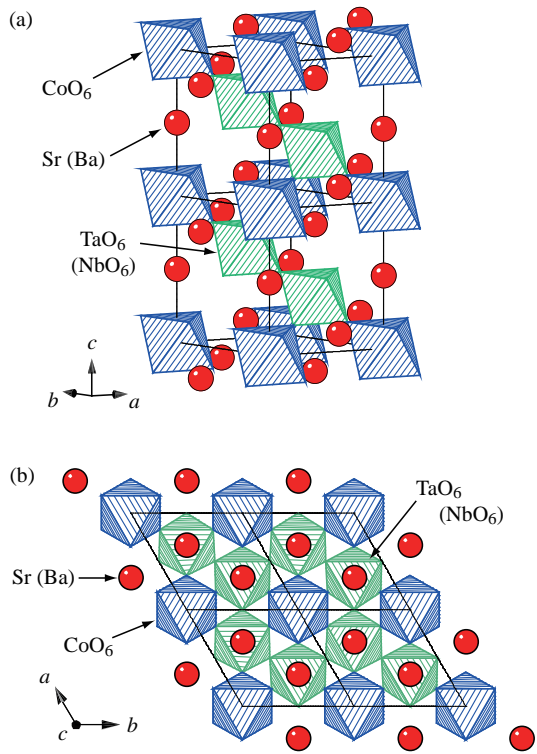}
	\caption{(Color online) (a) Schematic view of the crystal structure of Sr$_3$CoTa$_2$O$_9$ and related systems. Blue and green indicate single CoO$_6$ and TaO$_6$ (NbO$_6$) octahedra, respectively.  Solid lines denote the chemical unit cell. (b) Crystal structure viewed along the $c$ axis. Magnetic Co$^{2+}$ ions form a uniform triangular lattice in the $ab$ plane.}
	\label{fig:cryst}
\end{figure}

In this paper, we report the magnetic phase diagram of a spin--1/2 TLAF Sr$_3$CoTa$_2$O$_9$, which has a trigonal structure with the space group $P\bar{3}m1$ as shown in Fig.~\ref{fig:cryst}~\cite{Treiber_ZAAC1982,Treiber_JSSC1982}. The crystal structure is the same as those of Ba$_3$CoNb$_2$O$_9$~\cite{Treiber_ZAAC1982,Ting_JSSC2004} and Ba$_3$CoTa$_2$O$_9$~\cite{Treiber_ZAAC1982,Treiber_JSSC1982}. Magnetic Co$^{2+}$ ions with effective spin--1/2 in an octahedral environment~\cite{Abragam,Lines,Oguchi} form a uniform triangular lattice parallel to the crystallographic $ab$ plane.

Recently, the magnetic properties of Ba$_3$CoNb$_2$O$_9$~\cite{Lee_Nb,Yokota,Jiao} and Ba$_3$CoTa$_2$O$_9$~\cite{Lee,Ranjith} have been investigated by magnetization and specific heat measurements, and neutron scattering experiments. Both systems exhibit two-step antiferromagnetic phase transitions at $T_{\rm N1}\,{=}\,1.39~{\rm K}$ and $T_{\rm N2}\,{=}\,1.13~{\rm K}$ for Ba$_3$CoNb$_2$O$_9$~\cite{Yokota}, and $T_{\rm N1}\,{=}\,0.70~{\rm K}$ and $T_{\rm N2}\,{=}\,0.57~{\rm K}$ for Ba$_3$CoTa$_2$O$_9$~\cite{Ranjith} owing to the weak easy-axis anisotropy. The saturation fields were found to be ${\mu}_0H_{\rm s}\,{=}\,4.0$\,T for Ba$_3$CoNb$_2$O$_9$~\cite{Lee_Nb,Yokota} and ${\mu}_0H_{\rm s}\,{=}\,1.4\,{-}\,3.0$\,T for Ba$_3$CoTa$_2$O$_9$~\cite{Lee,Ranjith}. These saturation fields are much smaller than ${\mu_0}H_{\rm s}\,{\simeq}\,32$\,T for Ba$_3$CoSb$_2$O$_9$~\cite{Shirata,Susuki}, which approximates the 2D spin--1/2 Heisenberg TLAF. This finding indicates that the intralayer exchange interactions in Ba$_3$CoNb$_2$O$_9$ and Ba$_3$CoTa$_2$O$_9$ are much smaller than that in Ba$_3$CoSb$_2$O$_9$. 

The distinct difference between the exchange interactions in these systems can be ascribed to the filled outermost orbitals of nonmagnetic pentavalent ions. Superexchange interactions between neighboring spins in the same triangular layer occur through the Co$^{2+}${--}\,O$^{2-}${--}\,O$^{2-}${--}\,Co$^{2+}$ and Co$^{2+}${--}\,O$^{2-}${--}\,B$^{5+}${--}\,O$^{2-}${--}\,Co$^{2+}$ pathways. The superexchange through the former pathway should be antiferromagnetic, whereas the latter path leads to a ferromagnetic superexchange interaction for B$^{5+}$\,{=}\,Nb$^{5+}$ and Ta$^{5+}$ with the filled outermost $4p$ and $5p$ orbitals owing to the Hund rule in these orbitals~\cite{Yokota,Koga}. We consider that the superexchange interactions via these two pathways are almost canceled in the case of B$^{5+}$\,{=}\,Nb$^{5+}$ and Ta$^{5+}$, resulting in a weakly antiferromagnetic total exchange interaction. On the other hand, for B$^{5+}$\,{=}\,Sb$^{5+}$ with the filled outermost $4d$ orbital, the superexchange interaction through the Co$^{2+}${--}\,O$^{2-}${--}\,Sb$^{5+}${--}\,O$^{2-}${--}\,Co$^{2+}$ pathway becomes antiferromagnetic owing to the Pauli principle, and the total exchange interaction should be strongly antiferromagnetic~\cite{Yokota,Koga}. In this context, the exchange interaction in a triangular lattice of Sr$_3$CoTa$_2$O$_9$ is considered to be as small as those in Ba$_3$CoNb$_2$O$_9$ and Ba$_3$CoTa$_2$O$_9$. However, the magnitudes of the interlayer exchange interaction relative to the intralayer exchange interaction in these three systems will be different. To elucidate the change in the magnetic phase diagram arising from the difference in the relative magnitude between the intralayer and interlayer exchange interactions and between the anisotropy term and the interlayer exchange interaction, we investigated the low-temperature magnetic properties of Sr$_3$CoTa$_2$O$_9$ powder.

\section{Experimental details}

A powder sample of Sr$_3$CoTa$_2$O$_9$ was prepared by a solid-state reaction in accordance with the chemical reaction $3\mathrm{Sr}\mathrm{CO}_3+\mathrm{Co}\mathrm{O}+\mathrm{Ta}_2\mathrm{O}_5\longrightarrow \mathrm{Sr}_3\mathrm{Co}\mathrm{Ta}_2\mathrm{O}_9 + 3\mathrm{CO}_2$ in air. Reagent-grade materials were mixed at stoichiometric ratios and calcined at 1000\,$^\circ$C in the air for one day. After being pressed into a pellet, Sr$_3$CoTa$_2$O$_9$ was sintered four times at 1230, 1400, 1500, and 1600\,$^{\circ}$C for one day. With increasing sintering temperature, the sample color changed from yellowish-white to reddish-purple, and its quality was improved~\cite{Quality}.

The powder X--ray diffraction (XRD) measurement of Sr$_3$CoTa$_2$O$_9$ was conducted using a MiniFlex II diffractometer (Rigaku) with Cu $K\alpha$ radiation at room temperature. The crystal structure of Sr$_3$CoTa$_2$O$_9$ was refined by Rietveld analysis using the RIETAN--FP program~\cite{Izumi2007}.

Magnetization in the temperature range of $0.5\,{\leq}\,T\,{\leq}\,300$\,K in magnetic fields up to 7\,T was measured using a Magnetic Property Measurement System (MPMS--XL, Quantum Design) equipped with an iHelium3 option (IQUANTUM). Specific heat in the temperature range of $0.34\,{\leq}\,T\,{\leq}\,300$ K  in magnetic fields up to 9\,T was measured using a Physical Property Measurement System (PPMS, Quantum Design) by the relaxation method. Nuclear magnetic resonance (NMR) measurements were performed on a powder sample using a 16\,T superconducting magnet in the temperature range between 0.4 and 1.2\,K.

\section{Results and Discussion}\label{results}

\subsection{Crystal structure}\label{structure}
\begin{figure}[t]
	\centering
	\includegraphics[width=8.5cm]{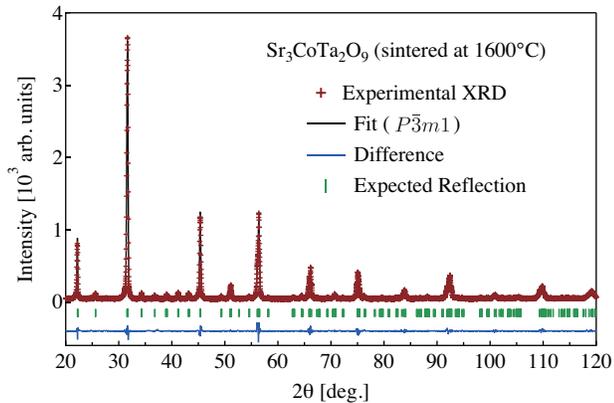}
	\caption{(Color online) XRD pattern of Sr$_3$CoTa$_2$O$_9$ measured at room temperature. Experimental data, the results of Rietveld fitting, their difference and expected reflections are shown by the red symbols, black line, blue line and vertical green bars, respectively.}
	\label{fig:XRD}
\end{figure}

Although the crystal structure of Sr$_3$CoTa$_2$O$_9$ was reported in Refs.~\cite{Treiber_ZAAC1982,Treiber_JSSC1982}, details of the structure parameters in Sr$_3$CoTa$_2$O$_9$ have not been reported. We refined the structure parameters by Rietveld analysis using RIETAN--FP~\cite{Izumi2007}.
The results of the XRD measurement of Sr$_3$CoTa$_2$O$_9$ at room temperature and the Rietveld analysis are shown in Fig. \ref{fig:XRD}. The analysis was based on the structural model with the space group $P\bar{3}m1$ as in Ba$_3$CoNb$_2$O$_9$~\cite{Treiber_ZAAC1982,Ting_JSSC2004} and Ba$_3$CoTa$_2$O$_9$~\cite{Treiber_ZAAC1982,Treiber_JSSC1982}. First, we chose the structure parameters of Ba$_3$CoTa$_2$O$_9$~\cite{Lee} as the initial parameters of the Rietveld analysis, setting the site occupancy to 1 for all atoms. 
The structure parameters refined for the space group $P\bar{3}m1$ using the XRD data are summarized in Table \ref{tab:Rietveld_XRD}. The lattice constants of Sr$_3$CoTa$_2$O$_9$ are $a\,{=}\,5.6439(2)$ {\AA} and $c\,{=}\,6.9533(3)$ {\AA}, which are smaller than $a\,{=}\,5.7737$ {\AA} and $c\,{=}\,7.0852$ {\AA} in Ba$_3$CoNb$_2$O$_9$~\cite{Ting_JSSC2004}, and $a\,{=}\,5.7748$ {\AA} and $c\,{=}\,7.0908$ {\AA} in Ba$_3$CoTa$_2$O$_9$~\cite{Ranjith}, because the ionic radius of Sr$^{2+}$ is smaller than that of Ba$^{2+}$. The CoO$_6$ octahedron in Sr$_3$CoTa$_2$O$_9$ displays a trigonal elongation along the $c$ axis, which gives rise to the weak easy-axis anisotropy~\cite{Lines}.

\begin{table}[t]
	\centering
	\caption{Structure parameters of Sr$_3$CoTa$_2$O$_9$ determined from XRD measurements at room temperature. $x, y$ and $z$ are the fractional atomic coordinates and $U$ is equivalent isotropic displacement parameter.}
	\label{tab:Rietveld_XRD}
	\begin{tabular}{clllll}
		\hline\hline
		Atom \ \ & Site & \ \ $x$ & \ \ $y$ & \ \ $z$ &\ \ $U$/{\AA}$^2$ \\\hline
		$\mathrm{Sr(1)}$\ \  & $2d$\ \   & 1/3 & 2/3 & 0.6606(4) &\ \ 0.0085(3) \\
		$\mathrm{Sr(2)}$\ \  & $1a$ & 0 & 0 & 0 &\ \ 0.0085(3) \\
		$\mathrm{Co}$ \ \ & $1b$\ \   & 0 & 0 & 1/2 &\ \ 0.0017(2)  \\
		$\mathrm{Ta}$ \ \ & $2d$ \ \  & 1/3 & 2/3 & 0.1755(3) &\ \ 0.0017(2) \\
		$\mathrm{O(1)}$\ \  & $3e$\ \  & 1/2 & 0 & 0 &\ \ 0.036(2) \\
		$\mathrm{O(2)}$\ \  & $6i$\ \  & 0.165(1)\ \  & 0.331(1) \ \ & 0.321(2) \ \ &\ \ 0.036(2) \\
		\hline\hline 
		\multicolumn{5}{c}{\vspace{0.3mm}Space group $P \bar{3} m 1$}\\
		\multicolumn{5}{c}{$a=5.6439(2)$ \AA, $c=6.9533(3)$ \AA}\\
		\multicolumn{5}{c}{$R_{\mathrm{wp}}=7.7\%$, $R_{\mathrm{e}}=11.0\%$.}\\\hline
	\end{tabular}
\end{table}

\subsection{Magnetic susceptibility and magnetization}

\begin{figure}[t]
	\centering
	\includegraphics[width=8.0cm]{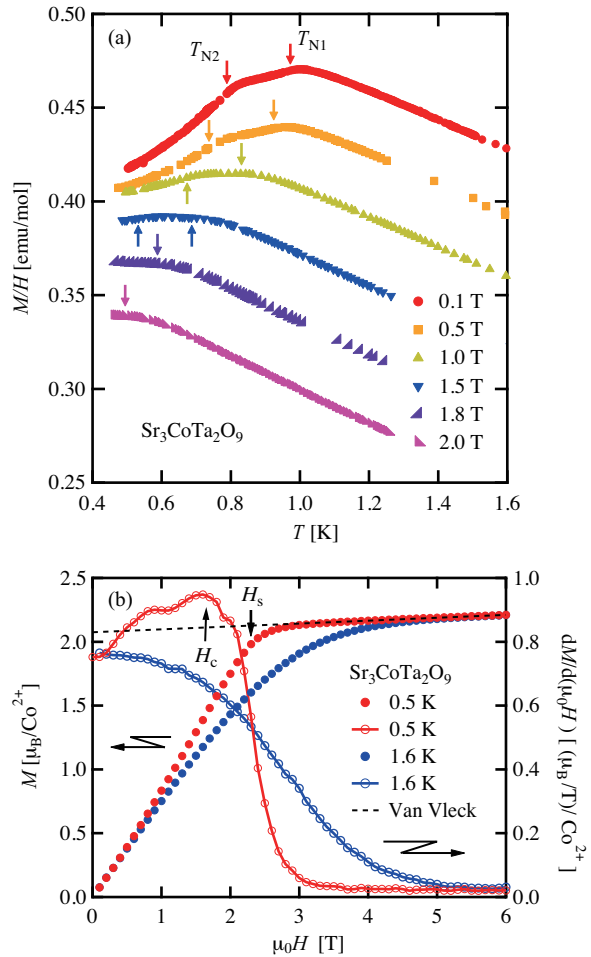}
	\caption{(Color online) (a) Temperature dependence of the magnetic susceptibility $M/H$ of Sr$_3$CoTa$_2$O$_9$ powder measured at various magnetic fields up to 2 T. Vertical arrows indicate the magnetic phase transition temperatures $T_{\rm N1}$ and $T_{\rm N2}$. (b) Field dependence of raw magnetization $M$ (left) and its field derivative $dM/d({\mu}_0H)$ (right) in Sr$_3$CoTa$_2$O$_9$ powder measured at 0.5 and 1.6\,K. The dashed line denotes the Van Vleck paramagnetism. The vertical arrows indicate the critical field $H_{\rm c}$ and saturation field $H_{\rm s}$ expected at $T\,{=}\,0$ K.}
	\label{fig:mag}
\end{figure}

\begin{figure}[t]
	\centering
	\includegraphics[width=8.0cm]{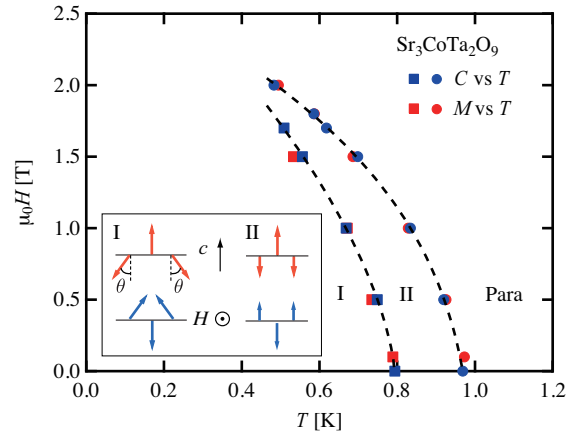}
	\caption{(Color online) $H\,{-}\,T$ phase diagram for Sr$_3$CoTa$_2$O$_9$ powder obtained from the present magnetization and specific heat measurements. Dashed curves are guides to the eye. The inset shows the spin configurations expected in the two ordered phases, I and II. Thick red and blue arrows denote the spin configurations on the two neighboring triangular layers parallel to the $ab$ plane. Para means the paramagnetic phase. The magnetic field is assumed to be perpendicular to the $c$ axis.}
	\label{fig:phase}
\end{figure}

Figure~\ref{fig:mag}(a) shows the temperature dependence of magnetization divided by the field $M/H$ in Sr$_3$CoTa$_2$O$_9$ measured at several magnetic fields of up to 2\,T. No thermal hysteresis is observed between the zero-field-cooled and field-cooled data. For ${\mu}_0H\,{=}\,0.1$\,T, two bend anomalies indicative of magnetic phase transitions are clearly observed at $T_{\rm N1}\,{=}\,0.97$\,K and $T_{\rm N2}\,{=}\,0.79$\,K. We assign these transition temperatures to the temperatures giving the peak of d$(M/H)$/d$T$, because these temperatures coincide with the temperatures giving peaks of the specific heat shown in the next subsection. With increasing magnetic field, $T_{\rm N1}$ and $T_{\rm N2}$ shift to the low-temperature side. The transition temperatures $T_{\rm N1}$ and $T_{\rm N2}$ obtained at various magnetic fields are summarized in a magnetic field vs temperature $(H\,{-}\,T)$ diagram shown in Fig.~\ref{fig:phase}. Because the magnetization anomalies at $T_{\rm N1}$ and $T_{\rm N2}$ become more smeared with increasing magnetic field, there is a certain error in determining these transition temperatures in magnetic fields.

Figure~\ref{fig:mag}(b) shows the raw magnetization $M$ and its field derivative $dM/d({\mu}_0H)$ as functions of ${\mu}_0H$ measured at $T\,{=}\,0.5$ and 1.6\,K. The saturation anomaly observed at $T\,{=}\,1.6$\,K is indistinct owing to the finite temperature effect, whereas that observed at $T\,{=}\,0.5$\,K is fairly sharp. The saturation field expected at $T\,{=}\,0$ K is estimated to be ${\mu}_0H_{\rm s}\,{=}\,2.3$\,T from the inflection point of the $dM/d({\mu}_0H)$ curve at $T\,{=}\,0.5$\,K. This saturation field is of the same order of magnitude as ${\mu}_0H_{\rm s}\,{=}\,4.0$\,T for Ba$_3$CoNb$_2$O$_9$~\cite{Lee_Nb,Yokota} and ${\mu}_0H_{\rm s}\,{=}\,1.4\,{-}\,3.0$\,T for Ba$_3$CoTa$_2$O$_9$~\cite{Lee,Ranjith}, but much smaller than ${\mu}_0H_{\rm s}\,{\simeq}\,32$\,T for Ba$_3$CoSb$_2$O$_9$~\cite{Shirata,Susuki}. The small saturation field in Sr$_3$CoTa$_2$O$_9$ demonstrates that its intralayer exchange interaction is much smaller than that in Ba$_3$CoSb$_2$O$_9$. This distinct difference in intralayer exchange interaction between Ta (Nb) and Sb systems can be ascribed to the filled outermost orbitals of nonmagnetic pentavalent ions, as mentioned in Sec.~\ref{Intro}. 

Above the saturation field $H_{\rm s}$, magnetization increases linearly with increasing field because of the large temperature-independent Van Vleck paramagnetism characteristic of the Co$^{2+}$ ion in the octahedral environment~\cite{Oguchi}. 
From the magnetization slope above $H_{\rm s}$, we evaluate the Van Vleck paramagnetic susceptibility to be
$\chi_{\rm VV}\,{=}\,2.3 \times 10^{-2}~\mu_{\rm B}/({\rm Co}^{2+}{\rm T})\,{=}\,1.3\,{\times}\,10^{-2}~{\rm emu/mol}$ in Sr$_3$CoTa$_2$O$_9$. By subtracting the Van Vleck term, we obtain the saturation magnetization to be $M_{\rm s}\,{=}\,2.08~{\mu_{\rm B}/{\rm Co}^{2+}}$, which gives an average $g$ factor of 4.15. This $g$ factor is significantly greater than $g\,{=}\,3.0$ in Ba$_3$CoNb$_2$O$_9$~\cite{Yokota} and $g\,{=}\,3.1\,{-}\,3.6$ in Ba$_3$CoTa$_2$O$_9$~\cite{Lee,Ranjith}. This indicates that the reduction in the orbital moment of Co$^{2+}$ due to the mixing of the $p$ orbital of ligand O$^{2-}$ is significantly small in Sr$_3$CoTa$_2$O$_9$, because the average $g$ factor for three different field directions without the orbital reduction is given by $g_{\rm avg}\,{\approx}\,4.3$~\cite{Abragam,Lines}.

As seen from Fig.\,\ref{fig:mag}(b), the magnetization anomaly at the saturation field $H_{\rm s}$ observed at $T\,{=}\,0.5$\,K is considerably sharp despite the sample being in powder form. If the $g$ factor and/or exchange interaction is anisotropic, as reported for many Co$^{2+}$ compounds, the saturation field depends strongly on the field direction. Consequently, the magnetization anomaly at $H_{\rm s}$ for powder samples should be smeared. The sharp magnetization anomaly at $H_{\rm s}$ observed in Sr$_3$CoTa$_2$O$_9$ demonstrates that both the $g$ factor and the exchange interaction are nearly isotropic, as in the cases of Ba$_3$CoSb$_2$O$_9$~\cite{Susuki} and Ba$_3$CoNb$_2$O$_9$~\cite{Lee_Nb,Yokota}.

The magnetic anisotropy in Sr$_3$CoTa$_2$O$_9$ is deduced to be of the easy-axis type because the observed two-step magnetic ordering is characteristic of TLAFs with the easy-axis anisotropy~\cite{Miyashita}. The easy-axis anisotropy is also anticipated from the trigonally elongated CoO$_6$ octahedron, as shown in \ref{structure}~\cite{Lines}. If the interlayer exchange interaction is ferromagnetic, the 1/3--plateau will emerge even in classical TLAFs when a magnetic field is applied parallel to the easy axis~\cite{Miyashita}. The classical 1/3--magnetization plateau has been observed in the quasi-2D large spin TLAFs GdPd$_2$Al$_3$~\cite{Kitazawa,Inami} and Rb$_4$Mn(MoO$_4$)$_3$~\cite{Ishii}. As shown in Fig.~\ref{fig:mag}(b), the magnetization curve displays no magnetization plateau at one-third of the saturation magnetization. This result indicates that the interlayer exchange interaction is antiferromagnetic and of the same order of magnitude as the intralayer exchange interaction because the 1/3-quantum-plateau emerges only for TLAFs with a relatively small interlayer exchange interaction~\cite{Gekht,Yamamoto2,Li}.

\subsection{Specific heat}

\begin{figure}[t]
\centering
	\includegraphics[width=8.0cm]{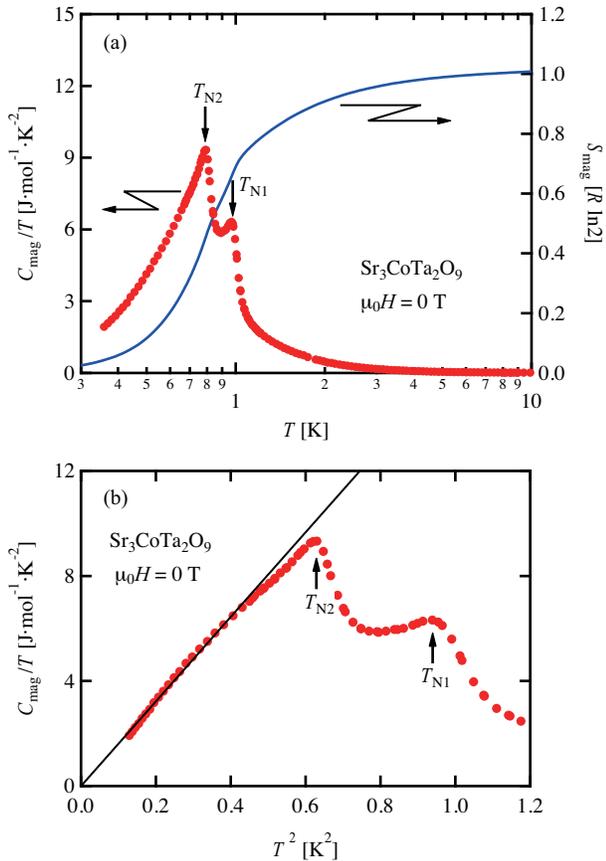}
\caption{(Color online) (a) Temperature dependence of magnetic specific heat divided by temperature $C_{\rm mag}/T$ in Sr$_3$CoTa$_2$O$_9$ measured at zero magnetic field. Magnetic phase transitions are observed at $T_{\rm N1}\,{=}\,0.97~{\rm K}$ and $T_{\rm N2}\,{=}\,0.79~{\rm K}$, as indicated by vertical arrows. The solid curve represents the magnetic entropy $S_{\rm mag}$ in units of $R\ln{2}$. (b) $C_{\rm mag}/T$ as a function of $T^2$. The solid line is a linear fit without a constant term.}
	\label{fig:heat_zero}
\end{figure}

Figure~\ref{fig:heat_zero}(a) shows the temperature dependence of magnetic specific heat in Sr$_3$CoTa$_2$O$_9$ divided by the temperature $C_{\rm mag}/T$ measured at zero magnetic field. The lattice contribution of the specific heat $C_{\rm latt}$ was evaluated by fitting a model function composed of the linear combination of the Debye model and Einstein mode to the experimental data between 10 and 32\,K, in which range the magnetic contribution is negligible. With decreasing temperature, $C_{\rm mag}/T$ increases gradually below 5 K and exhibits two pronounced peaks at $T_{\rm N1}\,{=}\,0.97$ K and  $T_{\rm N2}\,{=}\,0.79$ K indicative of two-step magnetic phase transitions. 
These transition temperatures are consistent with those identified by the magnetization measurements shown in Fig.~\ref{fig:mag}(a).

We evaluate the magnetic entropy $S_{\rm mag}$ of Sr$_3$CoTa$_2$O$_9$  by integrating the measured $C_{\rm mag}/T$ with respect to $T$ below 10 K. With increasing temperature, $S_{\rm mag}$ increases gradually and reaches almost $R\ln2$ at 10 K. This demonstrates that an effective spin--1/2 description is indeed valid for Sr$_3$CoTa$_2$O$_9$ in the low-temperature range of $T\,{<}\,10$ K.
Below $T_{\rm N2}$, the magnetic specific heat is proportional to $T^3$ with no residual term as shown in Fig.~\ref{fig:heat_zero}(b), which indicates that spin-wave dispersions are 3D in the ordered state.

\begin{figure}[t]
	\centering
	\includegraphics[width=8.0cm]{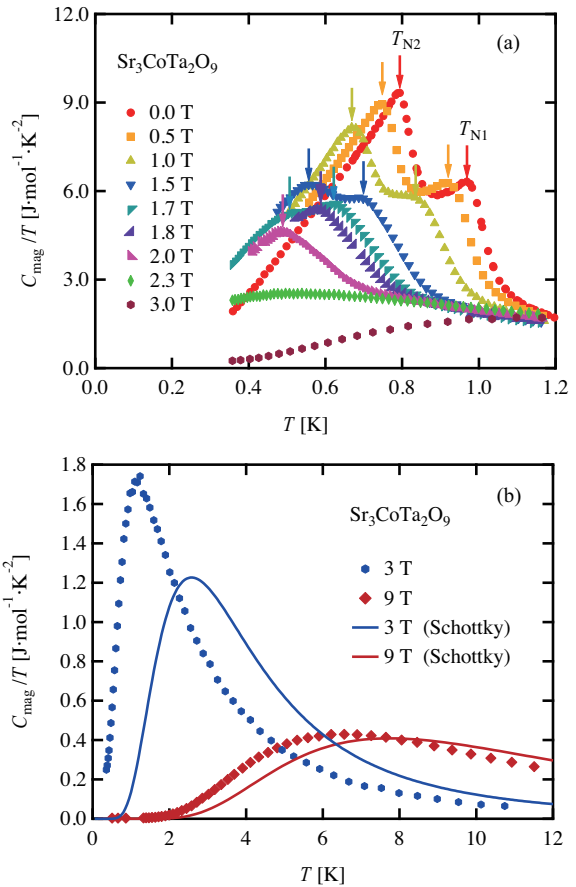}
	\caption{(Color online) Temperature dependence of $C_{\rm mag}/T$ in Sr$_3$CoTa$_2$O$_9$ measured at magnetic fields of (a) ${\mu}_0H\,{=}\,0-3$\,T for $T\,{<}\,1.2$\,K, and (b) 3 and 9\,T for $T\,{<}\,12$\,K. Solid curves in (b) represent the Schottky specific heats calculated for ${\mu}_0H\,{=}\,3$ and 9\,T.}
	\label{fig:heat_fields}
\end{figure}

Figure~\ref{fig:heat_fields}(a) shows $C_{\rm mag}/T$ vs $T$ measured at magnetic fields of up to 3\,T. As the magnetic field increases, the two peaks corresponding to $T_{\rm N1}$ and $T_{\rm N2}$ shift to the low-temperature side and broaden out. It is considered that the broadening of peaks in the present powder sample is due to the fact that the transition temperature depends on the field direction and its distribution becomes wider with increasing magnetic field. Specific heat anomalies associated with $T_{\rm N1}$ and $T_{\rm N2}$ are no longer detectable for ${\mu}_0H\,{>}\,2.3$\,T and ${\mu}_0H\,{>}\,1.8$\,T, respectively. The transition data are summarized in Fig.~\ref{fig:phase}. The transition points determined from the temperature dependences of magnetization and specific heat are consistent with each other.

At ${\mu}_0H\,{=}\,3$ and 9\,T, which are higher than the saturation field of ${\mu}_0H_{\rm s}\,{=}\,2.3$\,T, single broad peaks are observed at around $T_{\rm max}\,{=}\,1.2$ and 6.6\,K, respectively, as shown in Fig.~\ref{fig:heat_fields}(b). The broad peak at $T_{\rm max}$ shifts to higher temperatures with increasing magnetic field, whereas the peak height monotonically decreases. This behavior is characteristic of the Schottky-specific heat attributable to Zeeman splitting. The Schottky-specific heat of a spin-1/2 system without the exchange interaction between spins is expressed by
\begin{eqnarray}
C_{\rm Sch}=Nk_{\rm B}\left(\frac{g\mu_{\rm B}H}{2k_{\rm B}T}\right)^2\frac{1}{\cosh^2(g\mu_{\rm B}H/2k_{\rm B}T)},
\label{eq:schottky}
\end{eqnarray}
where $N$ is the number of atoms. Solid lines in Fig.~\ref{fig:heat_fields}(b) are Schottky-specific heats calculated for ${\mu}_0H\,{=}\,3$ and\,9 T with $g\,{=}\,4.15$. The calculated $T_{\rm max}$ is higher than the experimental $T_{\rm max}$, and the calculated peak height at $T_{\rm max}$ is smaller than the experimental peak height. This indicates that the gap in the sample is smaller than $g{\mu}_{\rm B}H$ because of the exchange interactions between spins, which produce a gapless ordered state until saturation. We see that the agreement between the experimental $C_{\rm mag}$ and $C_{\rm Sch}$ for ${\mu}_0H\,{=}\,9$\,T, which is considerably higher than the saturation field of ${\mu}_0H_{\rm s}\,{=}\,2.3$\,T, is much better than that for ${\mu}_0H\,{=}\,3$\,T, which is close to ${\mu}_0H_{\rm s}$.

\begin{figure}[t]
	\centering
	\includegraphics[width=7.5cm]{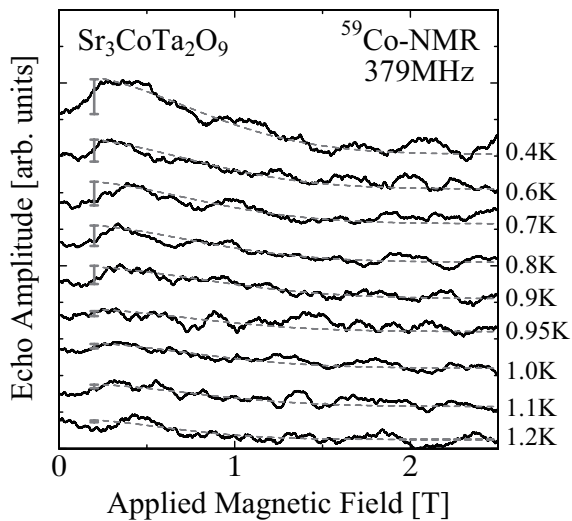}
	\caption{(Color online) $^{59}$Co-NMR spectra measured at various temperatures from 0.4 to 1.2\,K.  Dashed curves are guides to the eye, showing the horn-shaped powder pattern.  Vertical bars indicate the amplitude reduction from the top of the horn edge at 0.2 T to averaged value at around the zero field.}
	\label{fig:NMR_1}
\end{figure}

\subsection{NMR}
Figure~\ref{fig:NMR_1} shows the $^{59}$Co-NMR field-swept spectra observed under the low magnetic field region below 2.5\,T, measured with the resonance frequency ${\omega}/2{\pi}\,{=}\,379$\,MHz. The NMR spectra show a horn-shaped powder pattern with a broad tail toward higher fields. The horn edge locates at 0.2\,T, which is almost temperature independent. This type of spectral shape is known to appear for the powder sample in the magnetically ordered state, especially when the hyperfine field is close to ${\omega}/{\gamma}$, where ${\gamma}$ is the gyromagnetic ratio of $^{59}$Co nucleus~\cite{GR1}. If this condition is assumed to be fulfilled in the present case, we can roughly estimate the hyperfine field at the Co site to be 38(1)\,T, which is the quite typical value for Co-based oxide antiferromagnets, supporting the assumption~\cite{GR2,GR3}. We consider that the observed horn-shaped pattern is contributed from the central transition of $I\,{=}\,7/2$ $^{59}$Co nuclei with an electric quadrupole. The signal from the other transitions may be broadened due to the electric quadrupole interaction and become a background in a wide field region. The finite signal amplitude in the field region lower than 0.2\,T is explained by this contribution. We also refer to the finite signal amplitude above $T_{\rm N1}$, where the system is paramagnetic. This is simply due to the short-range magnetic order, possibly appearing slightly above $T_{\rm N1}$. Actually, the anomaly in $C_{\rm mag}$ starts from a temperature higher than $T_{\rm N1}$ by 0.1\,K, as shown in Fig.~\ref{fig:heat_fields}\,(a). 

\begin{figure}[t]
	\centering
	\includegraphics[width=8.0cm]{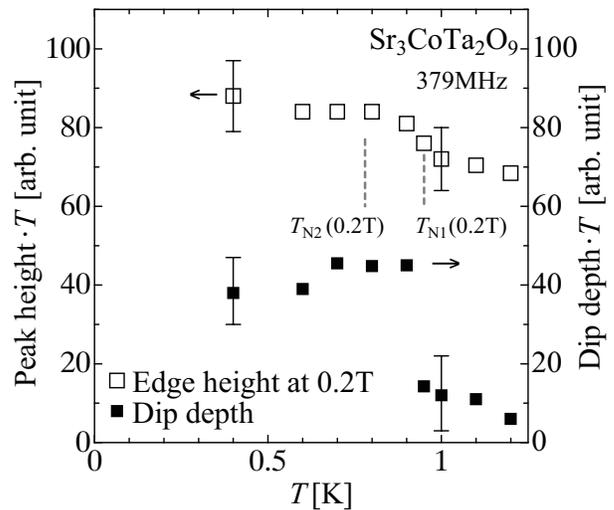}
	\caption{(Color online) Temperature dependence of the two parameters obtained from $^{59}$Co-NMR spectra: the horn edge height (open square) and the dip depth at the lower field side of the horn (solid square).  Error bars indicate typical uncertainty in determining each point.  See the text for the detailed definition of two quantities. Vertical dotted lines show $T_{\rm N1}$ or $T_{\rm N2}$ at 0.2\,T, estimated by the interpolation of data shown in Fig.~\ref{fig:phase}.}
	\label{fig:NMR_2}
\end{figure}

To extract information on the temperature dependence from the observed data, we focus attention on two points in the spectra in Fig.~\ref{fig:NMR_1}. One is the horn edge height at 0.2\,T, and the other is the dip depth from the top of the horn edge toward the averaged amplitude at around zero field. In Fig.~\ref{fig:NMR_2}, these two, multiplied by temperature, are plotted against the temperature. Note that both quantities do not show any marked temperature dependence below $T_{\rm N2}$. However, above $T_{\rm N2}$, they show slightly different behavior. That is, as increasing temperature, while the peak height (open symbol in Fig.~\ref{fig:NMR_2}) starts decreasing at $T_{\rm N2}$, the dip depth (closed symbol) stays constant at around $T_{\rm N2}$ and abruptly decreases at $T_{\rm N1}$. This difference supports the scenario of the two-step magnetic phase transition, as argued in the previous subsections.

Here, we discuss the change in the NMR spectrum at $T_{\rm N2}$. Generally, in the magnetically ordered state, the shape of the NMR powder pattern of the central transition line (${\pm}1/2$), as long as the electric quadrupole interaction is small, must be independent of the detail of the spin structure, that is, of the relative angle of neighboring spins. Thus, the observed change in the peak height (open symbol in Fig.~\ref{fig:NMR_2}) between $T_{\rm N2}$ and $T_{\rm N1}$ can be attributed to the change in the size of the hyperfine field at $T_{\rm N2}$, where the spin direction alters appreciably, as illustrated in the inset of Fig.~\ref{fig:phase}. Because the total hyperfine field at the Co site is the vector sum of onsite contribution and of supertransferred contribution, its size must change depending delicately on the relative angle of adjacent spins~\cite{GR4, GR5}.

\subsection{Phase diagram}

As shown in Fig.~\ref{fig:phase}, Sr$_3$CoTa$_2$O$_9$ undergoes two magnetic phase transitions at $T_{\rm N1}\,{=}\,0.97$ K and $T_{\rm N2}\,{=}\,0.79$ K at zero magnetic field, as similarly observed in Ba$_3$CoNb$_2$O$_9$~\cite{Lee_Nb,Yokota} and Ba$_3$CoTa$_2$O$_9$~\cite{Ranjith}. It has been theoretically demonstrated that two-step magnetic ordering occurs in TLAFs with the magnetic anisotropy of the easy-axis type, whereas the ordering is in a single-step in the case of the easy-plane anisotropy~\cite{Miyashita,Plumer_PRL1988,Plumer_PRB1989,Matsubara,MK}. 
The two-step magnetic phase transitions were observed in many hexagonal ABX$_3$-type TLAFs with the easy-axis anisotropy, in which triangular layers are coupled by strong antiferromagnetic exchange interactions along the $c$ axis ~\cite{Collins}. 
Thus, the successive phase transitions observed in Sr$_3$CoTa$_2$O$_9$ are attributed to the easy-axis anisotropy. There is no other reasonable description of the two-step magnetic ordering because the triangular lattice in Sr$_3$CoTa$_2$O$_9$ is uniform. The origin of the easy-axis anisotropy is not the single-ion anisotropy of the form $D(S_i^z)^2$ but the anisotropic exchange interaction of the form ${\Delta}JS_i^zS_j^z$ because the effective spin is $S\,{=}\,1/2$. The anisotropic exchange interaction of the easy-axis type is compatible with the trigonal elongation of the CoO$_6$ octahedron along the $c$ axis observed by the present structural refinement~\cite{Lines}. 

The scenario of the two-step magnetic ordering is as follows~\cite{Miyashita,Plumer_PRL1988,Plumer_PRB1989,Matsubara,MK}. With decreasing temperature, the $z$ components of spins order first at $T_{\rm N1}$ with the ferrimagnetic structure in a triangular layer. With further decreasing temperature, the $xy$ components of spins order at $T_{\rm N2}$, such that the spins form a triangular structure in the plane including the $c$ axis, as shown in the inset of Fig.~\ref{fig:phase}. The canting angle $\theta$ increases with decreasing temperature. Ordered spins on the same sites of the neighboring triangular layers should be antiparallel owing to the antiferromagnetic interlayer exchange interaction. If the interlayer exchange interaction is ferromagnetic, a weak total net moment will emerge, as observed in Ba$_2$La$_2$NiTe$_2$O$_{12}$~\cite{Saito}, because each triangular layer has a weak net moment along the $c$ axis. However, no total net moment was observed in the ordered state of Sr$_3$CoTa$_2$O$_9$, which indicates that net moments of neighboring layers cancel out due to the antiparallel spin arrangement along the $c$ axis.

The temperature range of the intermediate phase normalized by $T_{\rm N1}$ at zero magnetic field as $(T_{\rm N1}\,{-}\,T_{\rm N2})/T_{\rm N1}$ is a measure of the magnitude of the easy-axis anisotropy relative to the intralayer exchange interaction~\cite{Matsubara,Miyashita,MK}. In Sr$_3$CoTa$_2$O$_9$, $(T_{\rm N1}\,{-}\,T_{\rm N2})/T_{\rm N1}\,{=}\,0.18$. The narrow intermediate phase indicates that the easy-axis anisotropy is considerably smaller than the intralayer interaction. 
The absence of the 1/3--magnetization plateau also indicates that the interlayer exchange interaction is antiferromagnetic and its magnitude is comparable to that of the intralayer exchange interaction. Thus, Sr$_3$CoTa$_2$O$_9$ can be magnetically described as an antiferromagnetically stacked 3D TLAF with a small easy-axis anisotropy as Ba$_3$CoNb$_2$O$_9$~\cite{Lee_Nb,Yokota,Jiao} and Ba$_3$CoTa$_2$O$_9$~\cite{Lee,Ranjith}. Neutron diffraction experiments determined the propagation vector below $T_{\rm N2}$ in Ba$_3$CoNb$_2$O$_9$ as ${\bm k}\,{=}\,(1/3, 1/3,1)$~\cite{Lee_Nb,Jiao}, demonstrating the triangular spin ordering in the $ab$ plane and the antiferromagnetic ordering along the $c$ axis. 

\begin{figure}[t]
	\centering
	\includegraphics[width=5.5cm]{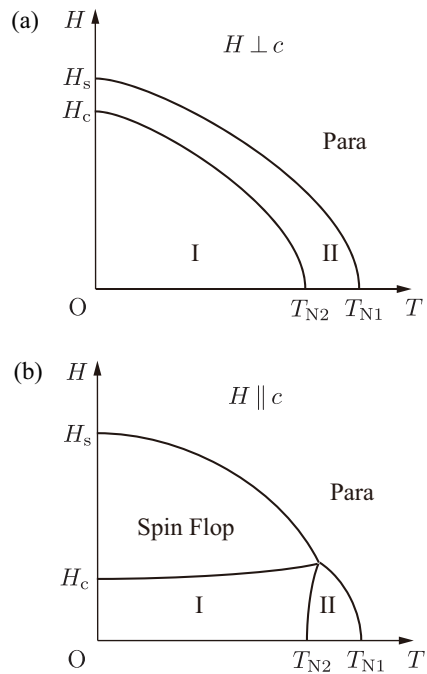}
	\caption{(Color online) Schematic phase diagrams of the 3D TLAF with easy-axis anisotropy for (a) $H\,{\perp}\,c$ and (b) $H\,{\parallel}\,c$~\cite{Plumer_PRL1988,Plumer_PRB1989}. The spin structure of phase I is illustrated in Figs.~\ref{fig:spin_config}\,(a) and (b) with ${\theta}\,{\neq}\,0$. The spin structure of phase II is given by ${\theta}\,{=}\,0$, i.e., for $H\,{\perp}\,c$, the spin components perpendicular to the magnetic field form a collinear structure along the $c$ axis. The spin structure of the spin flop phase is illustrated in Fig.~\ref{fig:spin_config}\,(c).}
	\label{fig:phase_schematic}
\end{figure}

\begin{figure*}[t]
	\centering
	\includegraphics[width=16.0cm]{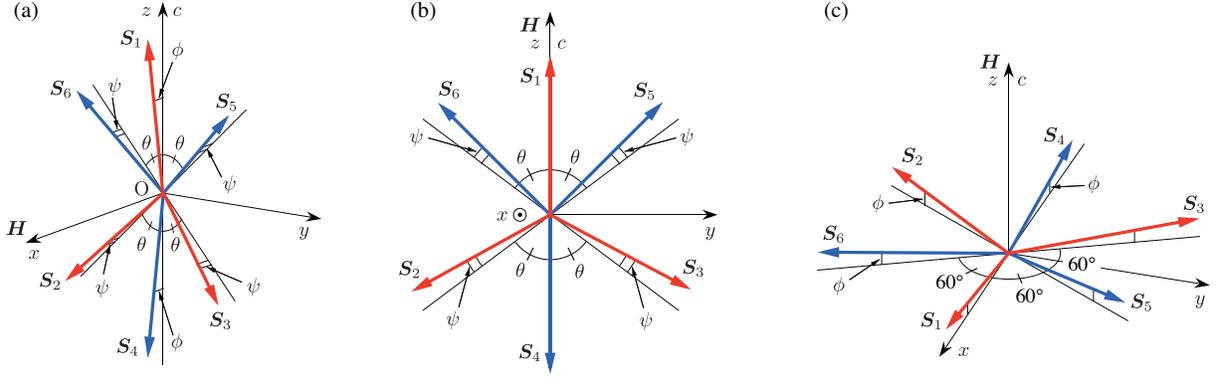}
	\caption{(Color online) Configurations of the sublattice spins for (a) $H\,{\perp}\,c$, (b) $H\,{\parallel}\,c$ and $H\,{<}\,H_{\rm c}$, and (c) $H\,{\parallel}\,c$ and $H\,{>}\,H_{\rm c}$ (spin flop phase). ${\bm S}_i$ $(i\,{=}\,1, 2$, and 3) and ${\bm S}_j$ $(j\,{=}\,4, 5$, and 6) are sublattice spins on odd-numbered and even-numbered triangular layers, respectively. The $z$ axis is taken to be parallel to the $c$ axis. In (a), the magnetic field is parallel to the $x$ axis, and $\theta$ is an angle between the $yz$ components of ${\bm S}_2$ (or ${\bm S}_3$) and the $c$ axis.}
	\label{fig:spin_config}
\end{figure*}

In the present experiment, we used a powder sample of Sr$_3$CoTa$_2$O$_9$. The transition anomaly for $H\,{\perp}\,c$ is more pronounced than that for $H\,{\parallel}\,c$, because the probability of $H\,{\perp}\,c$ is twice as large as that of $H\,{\parallel}\,c$ for the powder sample with the trigonal symmetry. Thus, we can deduce that the phase diagram shown in Fig.~\ref{fig:phase} approximates the phase diagram for $H\,{\perp}\,c$ in Sr$_3$CoTa$_2$O$_9$. This phase diagram is similar to that for$H\,{\perp}\,c$ in a 3D TLAF with easy-axis anisotropy, as shown in Fig.~\ref{fig:phase_schematic}\,(a), in which the antiferromagnetic interlayer exchange interaction is of the same order of magnitude as the intralayer exchange interaction, as discussed theoretically by Plumer {\it et al.}~\cite{Plumer_PRL1988,Plumer_PRB1989}. 

Phase boundaries for both $T_{\rm N1}$ and $T_{\rm N2}$ shift towards the low-temperature side with increasing magnetic field, as shown in Fig.~\ref{fig:phase}. We discuss the spin structure, assuming that the magnetic field is perpendicular to the $c$ axis for simplification. In this case, the spin components perpendicular to the magnetic field form a triangular structure in the low-temperature phase I, as shown in the inset of Fig.~\ref{fig:phase} and Fig.~\ref{fig:spin_config}\,(a). When the antiferromagnetic interlayer exchange interaction is comparable to the intralayer exchange interaction, the spin component parallel to the magnetic field increases with increasing magnetic field, and the canting angle $\theta$ decreases to zero at a critical field $H_{\rm c}$. This leads to a transition from phase I to phase II, in which spin components perpendicular to the magnetic field form a collinear structure along the $c$ axis~\cite{Plumer_PRB1989}.

\begin{figure}[t]
	\centering
	\includegraphics[width=8.0cm]{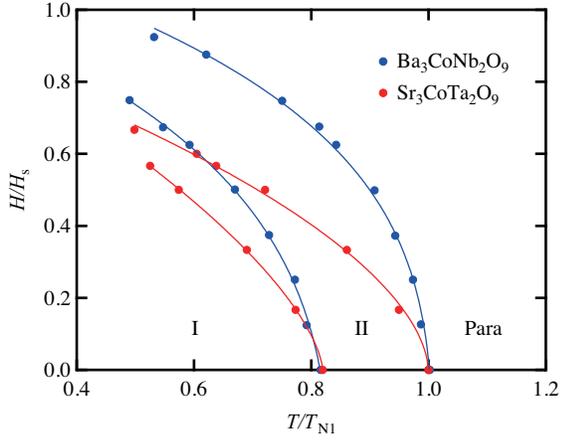}
	\caption{(Color online) $H\,{-}\,T$ phase diagrams in powder samples of Sr$_3$CoTa$_2$O$_9$ and Ba$_3$CoNb$_2$O$_9$~\cite{Yokota}, where temperature and magnetic field are normalized by the higher transition temperature $T_{\rm N1}$ at zero magnetic field and saturation field $H_{\rm s}$, respectively. Solid curves are guides to the eye. }
	\label{fig:phase_2}
\end{figure}

Figure~\ref{fig:phase_2} shows $H\,{-}\,T$ phase diagrams of powder samples of Sr$_3$CoTa$_2$O$_9$ and Ba$_3$CoNb$_2$O$_9$~\cite{Yokota}, where the temperature and magnetic field are normalized by the higher transition temperature $T_{\rm N1}$ at zero magnetic field and the saturation field $H_{\rm s}$, respectively. We did not show the normalized phase diagram of Ba$_3$CoTa$_2$O$_9$ in Fig.~\ref{fig:phase_2} because the saturation fields and phase diagrams reported in Refs.~\cite{Lee} and \cite{Ranjith} are significantly different. As mentioned previously, the phase diagram approximates the phase diagram for $H\,{\perp}\,c$. It is notable that the values of $T_{\rm N2}/T_{\rm N1}$ at zero magnetic field in these two systems are almost the same. This indicates that the ratios of the easy-axis anisotropy to the intralayer exchange interaction are almost the same in these systems. However, the phase boundaries of these systems behave differently in magnetic fields, although the phase boundaries shift towards the low-temperature side, as commonly observed with increasing magnetic field. 

To the author's knowledge, there is no detailed discussion of the phase transition between phases I and II for $H\,{\perp}\,c$ at $T\,{=}\,0$ in the 3D TLAF with the easy-axis anisotropy. Here, we discuss the phase transitions within the framework of the mean-field theory based on the six-sublattice model. We write the Hamiltonian in the magnetic field as
\begin{eqnarray}
{\cal H}\hspace{-1mm}&=&\hspace{-1mm}\sum_{\langle i,j\rangle} \left[J\left({\bm S}_i\cdot{\bm S}_j\right)+{\Delta}JS_i^zS_j^z\right]\nonumber\\
 \hspace{-1mm}&+&\hspace{-1mm} \sum_{\langle l,m\rangle} \left[J^{\prime}\left({\bm S}_l\cdot{\bm S}_m\right)+{\Delta}J^{\prime}S_l^zS_m^z\right]-\sum_i {\bm S}_i\cdot{\bm H},\hspace{3mm}
\label{eq:model}
\end{eqnarray}
where the first and second terms are the anisotropic exchange interactions in the triangular layer and between layers, respectively. For simplification, we put $g{\mu}_{\rm B}\,{=}\,1$. Details of the calculations are shown in Appendix. With increasing magnetic field for $H\,{\perp}\,c$, angles ${\phi}$ and ${\psi}$ increase, and angle ${\theta}$ decreases (see Fig.~\ref{fig:spin_config}\,(a)). The angle ${\theta}$ becomes zero at the critical $H_{\rm c}$, which is expressed as
\begin{widetext}
\begin{eqnarray}
H_{\rm c}=\sqrt{\frac{\left(6J+3{\Delta}J-2{\Delta}J^{\prime}\right)^2-9(J+{\Delta}J)^2}{\left(6J+3{\Delta}J-2{\Delta}J^{\prime}\right)^2(1+{\alpha})^2-9(J+{\Delta}J)^2}}\,\left\{9J+4J^{\prime}+2{\alpha}\left(3J+2J^{\prime}\right)\right\}S,
\label{eq:H_c}
\end{eqnarray}
\end{widetext}
with
\begin{eqnarray}
{\alpha}=\frac{\left(9J+6{\Delta}J-2{\Delta}J^{\prime}\right)\left(3{\Delta}J+2{\Delta}J^{\prime}\right)}{4J^{\prime}\left(6J+3{\Delta}J-2{\Delta}J^{\prime}\right)}.
\end{eqnarray}
The phase transition at $H_{\rm c}$ is the first order. We consider that an anomaly near 1.7 T in $dM/dH$ shown in Fig.~\ref{fig:mag}\,(b) corresponds to the critical field $H_{\rm c}$. 

Phase II above $H_{\rm c}$ is characterized by ${\theta}\,{=}\,0, {\phi}\,{\neq}\,90^{\circ}$, and ${\psi}\,{\neq}\,90^{\circ}$. The spin components perpendicular to the magnetic field form a collinear structure along the $c$ axis. With increasing magnetic field, angles ${\phi}$ and ${\psi}$ increase and both become 90$^{\circ}$ at the saturation field $H_{\rm s}$, which is given by
\begin{eqnarray}
H_{\rm s}=\left(9J+4J^{\prime}+3{\Delta}J+2{\Delta}J^{\prime}\right)S.
\label{eq:H_s}
\end{eqnarray}

The normalized magnetic-field range of phase II ${\Delta}h_{\rm II}\,{=}\,(H_{\rm s}\,{-}\,H_{\rm c})/H_{\rm s}$ increases with increasing ${\Delta}J/J$ and decreases with increasing $J^{\prime}/J$. In the 2D limit of $J^{\prime}\,{\rightarrow}\,0$, ${\Delta}J^{\prime}\,{\rightarrow}\,0$, and ${\alpha}\,{\rightarrow}\,{\infty}$, ${\Delta}h_{\rm II}$ is expressed as ${\Delta}h_{\rm II}\,{=}\,2J\sqrt{J(3J+2{\Delta}J)}/\{(2J+{\Delta}J)(3J+{\Delta}J)\}$. On the other hand, when $J^{\prime}\,{\gg}\,J, {\Delta}J$ and ${\Delta}J^{\prime}$ as in many hexagonal ABX$_3$ type TLAFs~\cite{Collins,Tanaka,Kambe}, ${\alpha}\,{\rightarrow}\,0$ and ${\Delta}h_{\rm II}\,{\rightarrow}\,0$. Extrapolating phase boundaries shown in Fig.~\ref{fig:phase_2} to $T\,{=}\,0$, we see that ${\Delta}h_{\rm II}$ for Sr$_3$CoTa$_2$O$_9$ is smaller than that for B$_3$CoNb$_2$O$_9$. Because the ratios of the easy-axis anisotropy to the intralayer exchange interaction are expected to be almost the same in these systems from almost the same $T_{\rm N2}/T_{\rm N1}$, we deduce that $J^{\prime}/J$ is greater in Sr$_3$CoTa$_2$O$_9$ than in B$_3$CoNb$_2$O$_9$.

\section{Summary}
We have reported the results of magnetization and specific heat measurements of Sr$_3$CoTa$_2$O$_9$ powder, in which Co$^{2+}$ ions with effective spin--1/2 form a uniform triangular lattice parallel to the $ab$ plane. It was found that the high sintering temperature of 1600$^{\circ}$C is necessary to obtain high-quality samples. The lattice parameters were refined by Rietveld analysis, as shown in Table \ref{tab:Rietveld_XRD}.

Sr$_3$CoTa$_2$O$_9$ exhibits two magnetic phase transitions at $T_{\rm N1}\,{=}\,0.97~{\rm K}$ and $T_{\rm N2}\,{=}\,0.79~{\rm K}$ at zero magnetic field, which arise from the competition between the intralayer exchange interaction and the weak easy-axis magnetic anisotropy. The magnetization saturates at ${\mu}_0H_{\rm s}\,{=}\,2.3$ T. From the saturation magnetization, the $g$ factor was evaluated to be $g\,{=}\,4.15$ on average. The magnetization curve displays no magnetization plateau at one-third of the saturation magnetization characteristic of quasi-2D quantum TLAFs. This indicates that the interlayer exchange interaction is antiferromagnetic and of the same order of magnitude as the intralayer exchange interaction. 

We obtained the $H\,{-}\,T$ phase diagram of Sr$_3$CoTa$_2$O$_9$ powder from magnetization and specific heat measurements in magnetic fields, which is considered to approximate that for $H\,{\perp}\,c$. This phase diagram is in accordance with that for a TLAF with a strong interlayer exchange interaction and a small easy-axis anisotropy~\cite{Plumer_PRL1988,Plumer_PRB1989}. Thus, Sr$_3$CoTa$_2$O$_9$ is characterized as a 3D spin--1/2 TLAF with a weak easy-axis anisotropy similar to Ba$_3$CoNb$_2$O$_9$~\cite{Lee_Nb,Yokota}. Based on the mean-field theory, we discussed phase transitions in the magnetic field for $H\,{\perp}\,c$ at $T\,{=}\,0$. We showed the critical field $H_{\rm c}$ at which the spin structure projected on a plane perpendicular to the magnetic field changes from the triangular structure (phase I) to the collinear structure along the $c$ axis (phase II). We observed a magnetization anomaly at $H_{\rm c}\,{\simeq}\,1.7$\,T suggestive of the transition between phases I and II. 

It was found from the $H\,{-}\,T$ phase diagrams of Sr$_3$CoTa$_2$O$_9$ and Ba$_3$CoNb$_2$O$_9$ normalized by $T_{\rm N1}$ and $H_{\rm s}$ that the values of $T_{\rm N2}/T_{\rm N1}$ are almost the same in these two systems. This indicates that the magnitudes of the easy-axis anisotropy relative to the intralayer exchange interaction are almost the same in these two systems. However, the normalized magnetic field range of phase II, ${\Delta}h_{\rm II}$, extrapolated to $T\,{=}\,0$, is smaller in Sr$_3$CoTa$_2$O$_9$ than in Ba$_3$CoNb$_2$O$_9$. This indicates that the ratio of the interlayer exchange interaction $J^{\prime}$ to the intralayer exchange interaction $J$ is larger in Sr$_3$CoTa$_2$O$_9$ than in Ba$_3$CoNb$_2$O$_9$. To elucidate the details of the phase diagram and magnetic parameters, experiments using single crystals should be carried out.

\appendix
\section{Phase transitions for $\bm {H\,{\perp}\,c}$}
Here, we derive the critical field $H_{\rm c}$ and saturation field $H_{\rm s}$ at $T\,{=}\,0$ for $H\,{\perp}\,c$ using the mean-field theory and a six-sublattice model shown in Fig.~\ref{fig:spin_config}. We follow the calculation process described in Refs.~\cite{Tanaka} and \cite{Kambe}. The energy per magnetic unit cell is given as
\begin{widetext}
\begin{eqnarray}
E\hspace{-1mm}&=&\hspace{-1mm}3J\left({\bm S}_1\,{\cdot}\,{\bm S}_2\,{+}\,{\bm S}_2\,{\cdot}\,{\bm S}_3\,{+}\,{\bm S}_3\,{\cdot}\,{\bm S}_1\,{+}\,{\bm S}_4\,{\cdot}\,{\bm S}_5\,{+}\,{\bm S}_5\,{\cdot}\,{\bm S}_6\,{+}\,{\bm S}_6\,{\cdot}\,{\bm S}_4\right)
+3{\Delta}J\left(S_1^zS_2^z\,{+}\,S_2^zS_3^z\,{+}\,S_3^zS_1^z\,{+}\,S_4^zS_5^z\,{+}\,S_5^zS_6^z\,{+}\,S_6^zS_4^z\right)\nonumber\\
\hspace{-1mm}&+&\hspace{-1mm}2J^{\prime}\left({\bm S}_1\cdot{\bm S}_4+{\bm S}_2\cdot{\bm S}_5+{\bm S}_3\cdot{\bm S}_6\right)+2{\Delta}J^{\prime}\left(S_1^zS_4^z+S_2^zS_5^z+S_3^zS_6^z\right)-\sum_{i=1}^6 {\bm S}_i\cdot{\bm H}.
\label{eq:energy}
\end{eqnarray}
Using angles ${\theta}, {\phi}$, and ${\psi}$ shown in Fig.~\ref{fig:spin_config}\,(a), sublattice spins ${\bm S}_i$ $(i\,{=}\,1, 2, \dots 6)$ are expressed as
\begin{eqnarray}
\left.
\begin{array}{l}
{\bm S}_1=S(\sin{\phi}, 0, \cos{\phi}),\hspace{2mm} {\bm S}_2=S(\sin{\psi}, -\sin{\theta}\cos{\psi}, -\cos{\theta}\cos{\psi}),\hspace{2mm} {\bm S}_3=S(\sin{\psi}, \sin{\theta}\cos{\psi}, -\cos{\theta}\cos{\psi}),\vspace{2mm}\\
{\bm S}_4=S(\sin{\phi}, 0, -\cos{\phi}),\hspace{2mm}
{\bm S}_5=S(\sin{\psi}, \sin{\theta}\cos{\psi}, \cos{\theta}\cos{\psi}),\hspace{2mm}
{\bm S}_6=S(\sin{\psi}, -\sin{\theta}\cos{\psi}, \cos{\theta}\cos{\psi}).
\end{array}
\right \} 
\label{eq:sublattice}
\end{eqnarray}
Substituting eq.~(\ref{eq:sublattice}) into eq.~(\ref{eq:energy}), we have
\begin{eqnarray}
E\hspace{-1mm}&=&\hspace{-1mm}-\,6J\left(-1+2\sin^2{\theta}\cos^2{\psi}-2\sin{\phi}\sin{\psi}+2\cos{\theta}\cos{\phi}\cos{\psi}\right)S^2-
6{\Delta}J\left(2\cos{\theta}\cos{\phi}\cos{\psi}-\cos^2{\theta}\cos^2{\psi}\right)S^2\nonumber\\
\hspace{-1mm}&-&\hspace{-1mm}2J^{\prime}\left(\cos2{\phi}+2\cos2{\psi}\right)S^2-2{\Delta}J^{\prime}\left(\cos^2{\phi}+2\cos^2{\psi}\right)S^2-2H(\sin{\phi}+2\sin{psi})S.
\end{eqnarray}
Equilibrium conditions are given by ${\partial}E/{\partial}{\theta}=0$, ${\partial}E/{\partial}{\phi}=0$, and ${\partial}E/{\partial}{\psi}=0$, which lead to
\begin{equation}
\left\{(6J+3{\Delta}J-2{\Delta}J^{\prime})\cos{\theta}\cos{\psi}-3(J+{\Delta}J)\cos{\phi}\right\}\sin{\theta}\cos{\psi}=0,
\end{equation}
\begin{equation}
H\cos{\phi}-6JS\cos{\phi}\sin{\psi}-6(J+{\Delta}J)S\cos{\theta}\sin{\phi}\cos{\psi}-2(2J^{\prime}+{\Delta}J^{\prime})S\sin{\phi}\cos{\phi}=0,
\end{equation}
and
\begin{eqnarray}
H\cos{\psi}\hspace{-1mm}&-&\hspace{-1mm}3(2J+{\Delta}J)S\sin^2{\theta}\sin{\psi}\cos{\psi}-3JS\sin{\phi}\cos{\psi}+{\Delta}JS\sin{\psi}\cos{\psi}-3(J+{\Delta}J)S\cos{\theta}\cos{\phi}\sin{\psi}\nonumber\\
\hspace{-1mm}&-&\hspace{-1mm}2J^{\prime}S\sin2{\psi}-{\Delta}J^{\prime}S\cos^2{\theta}\sin{\psi}\cos{\psi}=0.
\end{eqnarray}
\end{widetext}

In phase I, where ${\theta}\,{\neq}\,0$, the angle $\theta$ is expressed as
\begin{equation}
\cos{\theta}=\frac{3(J+{\Delta}J)}{6J+3{\Delta}J-2{\Delta}J^{\prime}}\,\frac{\cos{\phi}}{\cos{\psi}}.
\label{eq:theta}
\end{equation}
From eqs.~(A5)$-$(A7), we have
\begin{equation}
\sin{\phi}=\frac{H}{9J+4J^{\prime}+2{\alpha}(3J+2J^{\prime})},
\end{equation}
\begin{equation}
\sin{\psi}=\frac{(1+{\alpha})H}{9J+4J^{\prime}+2{\alpha}(3J+2J^{\prime})},
\end{equation}
where $\alpha$ is given by eq.~(4). The angle $\theta$ in eq.~(A7) becomes zero at the critical field $H_{\rm c}$. From this condition, we obtain the critical field expressed as eq.~(3).

In phase II, ${\theta}\,{=}\,0, {\phi}\,{\neq}\,90^{\circ}$, and ${\psi}\,{\neq}\,90^{\circ}$. Angles $\phi$ and $\psi$ become $90^{\circ}$ at the saturation field $H_{\rm s}$. When a condition $(\cos{\psi}/\cos{\phi})\,{\rightarrow}\,1/2$ is satisfied at $H_{\rm s}$, eqs.~(A5) and (A6) are equivalent. From this condition, we obtain the saturation field given by eq.~(5).

\section*{Acknowledgments}
We thank M. Watanabe for experimental support. This work was supported by Grants-in-Aid for Scientific Research (A) (No.~17H01142) and (C) (No.~19K03711) from Japan Society for the Promotion of Science.


\begin{thebibliography}{99} 

\bibitem{Collins} M. F. Collins and O. A. Petrenko, Triangular antiferromagnets, Can. J. Phys. {\bf 75}, 605 (1997).

\bibitem{Balents} L. Balents, Spin liquids in frustrated magnets, Nature \textbf{464}, 199 (2010).

\bibitem{Starykh2} O. A. Starykh, Unusual ordered phases of highly frustrated magnets: a review, Rep. Prog. Phys. \textbf{78}, 052502 (2015).

\bibitem{Miyashita} S. Miyashita, Magnetic properties of Ising-like Heisenberg antiferromagnets on the triangular lattice, J. Phys. Soc. Jpn. \textbf{55}, 3605 (1986).

\bibitem{Gekht} R. S. Gekht and I. N. Bondarenko, Triangular antiferromagnets with a layered structure in a uniform field, J. Exp. Theor. Phys. \textbf{84}, 345 (1997).

\bibitem{Nishimori} H. Nishimori and S. Miyashita, Magnetization process of the spin--1/2 antiferromagnetic Ising-like Heisenberg model on the triangular lattice, J. Phys. Soc. Jpn. {\bf 55} 4448 (1986).

\bibitem{Chubokov} A. V. Chubokov and D. I. Golosov, Quantum theory of an antiferromagnet on a triangular lattice in a magnetic field, J. Phys.: Condens. Matter \textbf{3}, 69 (1991).

\bibitem{Nikuni} T. Nikuni and H. Shiba, Quantum fluctuations and magnetic structures of CsCuCl$_3$ in high magnetic fields, J. Phys. Soc. Jpn. \textbf{62}, 3268 (1993).

\bibitem{Honecker} A. Honecker, A comparative study of the magnetization process of two-dimensional antiferromagnets, J. Phys.: Condens. Matter \textbf{11}, 4697 (1999).

\bibitem{Alicea} J. Alicea, A. V. Chubokov, and O. A. Starykh, Quantum stabilization of the 1/3-magnetization plateau in Cs$_2$CuBr$_4$, Phys. Rev. Lett. \textbf{102}, 137201 (2009).

\bibitem{Farnell} D. J. J. Farnell, R. Zinke, J. Schulenburg, and J. Richter, High-order coupled cluster method study of frustrated and unfrustrated quantum magnets in external magnetic fields, J. Phys.: Condens. Matter \textbf{21}, 406002 (2009).

\bibitem{Sakai} T. Sakai and H. Nakano, Critical magnetization behavior of the triangular- and kagome-lattice quantum antiferromagnets, Phys. Rev. B \textbf{83}, 100405(R) (2011).

\bibitem{Hotta}  C. Hotta, S. Nishimoto, and N. Shibata, Grand canonical finite size numerical approaches in one and two dimensions: real space energy renormalization and edge state generation, Phys. Rev. B \textbf{87}, 115128 (2013).

\bibitem{Yamamoto1} D. Yamamoto, G. Marmorini, and I. Danshita, Quantum phase diagram of the triangular-lattice XXZ model in a magnetic field, Phys. Rev. Lett. \textbf{112}, 127203 (2014).

\bibitem{Sellmann} D. Sellmann, X. F. Zhang, and S. Eggert, Phase diagram of the antiferromagnetic XXZ model on the triangular lattice, Phys. Rev. B \textbf{91}, 081104(R) (2015).

\bibitem{Coletta} T. Coletta, T. A. T\'{o}th, K. Penc, and F. Mila, Semiclassical theory of the magnetization process of the triangular lattice Heisenberg model, Phys. Rev. B \textbf{94}, 075136 (2016).

\bibitem{Kawamura} H. Kawamura and S. Miyashita, Phase transition of the Heisenberg antiferromagnet on the triangular lattice in a magnetic field, J. Phys. Soc. Jpn. \textbf{54}, 4530 (1985).

\bibitem{Lee_theory} D. H. Lee, J. D. Joannopoulos, J. W. Negele, and D. P. Landau, Symmetry analysis and Monte Carlo study of a frustrated antiferromagnetic planar (XY) model in two dimensions, Phys. Rev. B \textbf{33}, 450 (1986).

\bibitem{Seabra} L. Seabra, T. Momoi, P. Sindzingre, and N. Shannon, Phase diagram of the classical Heisenberg antiferromagnet on a triangular lattice in an applied magnetic field, Phys. Rev. B \textbf{84}, 214418 (2011).

\bibitem{Gvozdikova} M. V. Gvozdikova, P.-E. Melchy, and M. E. Zhitomirsky, Magnetic phase diagrams of classical triangular and kagome antiferromagnets, J. Phys.: Condens. Matter  \textbf{23}, 164209 (2011).

\bibitem{Nojiri_CsCuCl3} H. Nojiri, Y. Tokunaga, and M. Motokawa, Magnetic Phase Transition of Helical CsCuCl$_3$ in Magnetic Field, J. Phys. (Paris) {\bf 49}, Suppl. C8, 1459 (1988).

\bibitem{Sera} A. Sera, Y. Kousaka, J. Akimitsu, M. Sera, and K. Inoue, Pressure-induced quantum phase transitions in the $S=\frac{1}{2}$ triangular lattice antiferromagnet CsCuCl$_3$, Phys. Rev. B \textbf{96}, 014419 (2017).

\bibitem{Tazuke} Y. Tazuke, H. Tanaka, K. Iio, and K. Nagata, Magnetic susceptibility study of CsCuCl$_3$, J. Phys. Soc. Jpn. {\bf 50}, 3919 (1981).

\bibitem{Tanaka_JPSJ1992} H. Tanaka, U. Schotte, and K. D. Schotte, ESR Modes in CsCuCl$_3$, J. Phys. Soc. Jpn. {\bf 61}, 1344 (1992).

\bibitem{Ono1} T. Ono, H. Tanaka, H. Aruga Katori, F. Ishikawa, H. Mitamura, and T. Goto, Magnetization plateau in the frustrated quantum spin system Cs$_2$CuBr$_4$, Phys. Rev. B \textbf{67}, 104431 (2003).

\bibitem{Ono2} T. Ono, H. Tanaka, O. Kolomiyets, H. Mitamura, T. Goto, K. Nakajima, A. Oosawa, Y. Koike, K. Kakurai, J. Klenke, P. Smeibidle, and M. Mei{\ss}ner, Magnetization plateaux of the $S = 1/2$ two-dimensional frustrated antiferromagnet Cs$_2$CuBr$_4$, J. Phys.: Condens. Matter \textbf{16}, S773 (2004). 

\bibitem{Fortune} N. A. Fortune, S. T. Hannahs, Y. Yoshida, T. E. Sherline, T. Ono, H. Tanaka, and Y. Takano, Cascade of magnetic-field-induced quantum phase transitions in a spin--$\frac{1}{2}$ triangular-lattice antiferromagnet, Phys. Rev. Lett. \textbf{102}, 257201 (2009).

\bibitem{Shirata} Y. Shirata, H. Tanaka, A. Matsuo, and K. Kindo, Experimental realization of a spin--1/2 triangular-lattice Heisenberg antiferromagnet, Phys. Rev. Lett. \textbf{108}, 057205 (2012).

\bibitem{Zhou} H. D. Zhou, C. Xu, A. M. Hallas, H. J. Silverstein, C. R. Wiebe, I. Umegaki, J. Q. Yan, T. P. Murphy, J.-H. Park, Y. Qiu, J. R. D. Copley, J. S. Gardner, and Y. Takano, Successive phase transitions and extended spin-excitation continuum in the $S=1/2$ triangular-lattice antiferromagnet Ba$_3$CoSb$_2$O$_9$, Phys. Rev. Lett. \textbf{109}, 267206 (2012).

\bibitem{Susuki} T. Susuki, N. Kurita, T. Tanaka, H. Nojiri, A. Matsuo, K. Kindo, and H. Tanaka, Magnetization process and collective excitations in the $S=1/2$ triangular-lattice Heisenberg antiferromagnet Ba$_3$CoSb$_2$O$_9$, Phys. Rev. Lett. \textbf{110}, 267201 (2013).

\bibitem{Quirion} G. Quirion, M. Lapointe-Major, M. Poirier, J. A. Quilliam, Z. L. Dun, and H. D. Zhou, Magnetic phase diagram of Ba$_3$CoSb$_2$O$_9$ as determined by ultrasound velocity measurements, Phys. Rev. B \textbf{92}, 014414 (2015).

\bibitem{Koutroulakis} G. Koutroulakis, T. Zhou, Y. Kamiya, J. D. Thompson, H. D. Zhou, C. D. Batista, and S. E. Brown, Quantum phase diagram of the $S=\frac{1}{2}$ triangular-lattice antiferromagnet Ba$_3$CoSb$_2$O$_9$, Phys. Rev. B \textbf{91}, 024410 (2015).

\bibitem{Yamamoto2} D. Yamamoto, G. Marmorini, and I. Danshita, Microscopic model calculations for the magnetization process of layered triangular-lattice quantum antiferromagnets, Phys. Rev. Lett. \textbf{114}, 027201 (2015).

\bibitem{Okada} K. Okada, H. Tanaka, N. Kurita, D.Yamamoto, A. Matsuo, and K. Kindo, Field-orientation dependence of quantum phase transitions in the $S\,{=}\,\frac{1}{2}$ triangular-lattice antiferromagnet Ba$_3$CoSb$_2$O$_9$, Phys. Rev. B \textbf{106}, 104415 (2022).

\bibitem{Li} M. Li, M. L. Plumer, and G. Quirion, Effects of interlayer and bi-quadratic exchange coupling on layered triangular lattice antiferromagnets, J. Phys.: Condens. Matter \textbf{32}, 135803 (2020).

\bibitem{Plumer_PRL1988}
M. L. Plumer, K. Hood, and A. Caille, Multicritical point in the magnetic phase diagram of CsNiCl$_3$, Phys. Rev. Lett. \textbf{60}, 45 (1988).

\bibitem{Plumer_PRB1989}
M. L. Plumer, A. Caille, and K. Hood, Mnlticritical points in the magnetic phase diagrams of axial and planar antiferromagnets, Phys. Rev. \textbf{B 39}, 4489 (1989).

\bibitem{Treiber_ZAAC1982}
U. Treiber and S. Kemmler-Sack, {\" U}ber die Kationenordnung in Perowskiten mit hexagonaler BaTiO$_3$-Struktur vom Typ Ba3B$^{II}$Sb2O9 (B$^{II}$: Mg, Mn, Co, Ni, Cu, Zn), Z. Anorg. Allg. Chem. {\bf 487}, 161 (1982).

\bibitem{Treiber_JSSC1982}
U. Treiber and S. Kemmler-Sack, {\" U}ber Ordnung-Unordnungsph{\" a}nomene bei Sauerstoff perowskiten vom Typ A$_3^{2+}$B$^{2+}$M$_2^{5+}$O$_9$, J. Solid State Chem. \textbf{43}, 51 (1982).

\bibitem{Ting_JSSC2004}
V. Ting, Y. Liu, L. Nor\'en, R. L. Withers, D. J. Goossens, M. James, C. Ferraris, A structure, conductivity and dielectric properties investigation of A$_3$CoNb$_2$O$_9$ (A=Ca$^{2+}$, Sr$^{2+}$, Ba$^{2+}$) triple perovskites, J. Solid State Chem. {\bf 177}, 4428 (2004).

\bibitem{Abragam} A. Abragam and M. H. L. Pryce, The theory of paramagnetic resonance in hydrated cobalt salts, Proc. R. Soc. (London) A \textbf{206}, 173 (1951).

\bibitem{Lines} M. E. Lines, Magnetic properties of CoCl$_2$ and NiCl$_2$, Phys. Rev. \textbf{131}, 546 (1963).

\bibitem{Oguchi} T. Oguchi, Theory of magnetism in CoCl$_2$$\cdot$2H$_2$O, J. Phys. Soc. Jpn. \textbf{20}, 2236 (1965).

\bibitem{Lee_Nb} M. Lee, J. Hwang, E. S. Choi, J. Ma, C. R. Dela Cruz, M. Zhu, X. Ke, Z. L. Dun, and H. D. Zhou, Series of phase transitions and multiferroicity in the quasi-two-dimensional spin--$\frac{1}{2}$ triangular-lattice antiferromagnet Ba$_3$CoNb$_2$O$_9$, Phys. Rev. B \textbf{89}, 104420 (2014).

\bibitem{Yokota} K. Yokota, N. Kurita, and H. Tanaka, Magnetic phase diagram of the $S = 1/2$ triangular-lattice Heisenberg antiferromagnet Ba$_3$CoNb$_2$O$_9$, Phys. Rev. B \textbf{90}, 014403 (2014).

\bibitem{Jiao} J. Jiao, S. Zhang, Q. Huang, M. Zhang, M. F. Shu, G. T. Lin, C. R. dela Cruz, V. O. Garlea, N. Butch, M. Matsuda, H. Zhou, and J. Ma, Quantum Effect on the Ground State of the Triple-Perovskite Ba$_3$MNb$_2$O$_9$ (M\,{=}\,Co, Ni, and Mn) with Triangular-Lattice, Chem. Mater. \textbf{34}, 6617 (2022).

\bibitem{Lee} M. Lee, E. S. Choi, J. Ma, R. Sinclair, C. R. Dela Cruz, and H. D. Zhou., Magnetic and electric properties of triangular lattice antiferromagnets Ba$_3$ATa$_2$O$_9$ (A=Ni and Co), Mater. Res. Bull. \textbf{88}, 308 (2017).

\bibitem{Ranjith} K. M. Ranjith, K. Brinda, U. Arjun, N. G. Hegde, and R. Nath, Double phase transition in the triangular antiferromagnet Ba$_3$CoTa$_2$O$_9$, J. Phys.: Condens. Matter \textbf{29}, 115804 (2017).

\bibitem{Koga} T. Koga, N. Kurita, M. Avdeev, S. Danilkin, T. J. Sato, and H. Tanaka, Magnetic structure of the $S = \frac{1}{2}$ quasi-two-dimensional square-lattice Heisenberg antiferromagnet Sr$_2$CuTeO$_6$, Phys. Rev. B \textbf{93}, 054426 (2016).

\bibitem{Quality} Sr$_3$CoTa$_2$O$_9$ sample sintered at 1230$^{\circ}$C exhibits no specific heat anomalies at magnetic phase transition temperatures, while the sample sintered at 1600$^{\circ}$C displays sharp peak anomalies at $T_{\rm N1}\,{=}\,0.97$~K and $T_{\rm N2}\,{=}\,0.79$~K as shown in Fig.~\ref{fig:heat_zero}.

\bibitem{Izumi2007} F. Izumi and K. Momma, Three-dimensional visualization in powder diffraction, Solid State Phenom. \textbf{130}, 15 (2007).


\bibitem{Kitazawa} H. Kitazawa, H. Suzuki, H. Abe, J. Tang, and G. Kido, High-field magnetization of triangular lattice antiferromagnet: GdPd$_2$Al$_3$, Physica B \textbf{259-261}, 890 (1999).

\bibitem{Inami} T. Inami, N. Terada, H. Kitazawa, and O. Sakai, Resonant magnetic X-ray diffraction study on the triangular lattice antiferromagnet GdPd$_2$Al$_3$, J. Phys. Soc. Jpn. \textbf{78}, 084713 (2009).

\bibitem{Ishii} R. Ishii, S. Tanaka, K. Onuma, Y. Nambu, M. Tokunaga, T. Sakakibara, N. Kawashima, Y. Maeno, C. Broholm, D. P. Gautreaux, J. Y. Chan, and S. Nakatsuji, Successive phase transitions and phase diagrams for the quasi-two-dimensional easy-axis triangular antiferromagnet Rb$_4$Mn(MoO$_4$)$_3$, Europhys. Lett. \textbf{94}, 17001 (2011).

\bibitem{GR1} Y. Yamada and A. Sakata, An analysis method of antiferromagnetic powder patterns in spin-echo NMR under external fields, J. Phys. Soc. Jpn. \textbf{55}, 1751, (1986).

\bibitem{GR2} K. Okada and H. Yasuoka, $^{59}$Co NMR in an antiferromagnetic CoO single crystal, J. Phys. Soc. Jpn. \textbf{43}, 34 (1977).

\bibitem{GR3} T. Kubo, K. Adachi, M. Mekata, A. Hirai, $^{59}$Co nuclear magnetic resonance in antiferromagnetic CsCoCl$_3$, Solid State Commun. \textbf{29}, 553 (1979).

\bibitem{GR4} T. Goto, S. Nakajima, M. Kikuchi, Y. Syono, and T. Fukase, $^{63/65}$Cu/$^{203/205}$Tl NMR study on the antiferromagnetic phase of the Tl-based high-$T_{\rm c}$ oxide TlBa$_2$YCu$_2$O$_7$, Phys. Rev. B \textbf{54}, 3562 (1996).

\bibitem{GR5} T. Shimizu, T. Matsumoto, A. Goto, K. Yoshimura, and K. Kosuge, Cu\,{--}\,O\,{--}\,Cu bond-angle dependence of magnetic interactions in antiferromagnetic cuprates, Physica B \textbf{329-333}, 765 (2003).

\bibitem{Matsubara} F. Matsubara, Magnetic ordering in a hexagonal antiferromagnet, J. Phys. Soc. Jpn. \textbf{51}, 2424 (1982).

\bibitem{MK} S. Miyashita and H. Kawamura, Phase transition of anisotropic Heisenberg antiferromagnets on triangular lattice, J. Phys. Soc. Jpn. \textbf{54}, 3385 (1985).

\bibitem{Saito} M. Saito, M. Watanabe, N. Kurita, A. Matsuo, K. Kindo, M. Avdeev, H. O. Jeschke, and H. Tanaka, Successive phase transitions and magnetization plateau in the spin-1 triangular-lattice antiferromagnet, Phys. Rev. B \textbf{100},  064417 (2019).

\bibitem{Tanaka} H. Tanaka, S. Teraoka, E. Kakehashi, K. Iio, and K. Nagata, ESR in hexagonal ABX3-Type antiferromagnets: I. Ground state properties in easy-axis anisotropy case, J. Phys. Soc. Jpn.  \textbf{57}, 3979 (1988).

\bibitem{Kambe} T. Kambe, H. Tanaka, S. Kimura, H. Ohta, M. Motokawa, and K. Nagata, Effects of diagonal interchain exchange interaction on the ESR modes of ABX3-type triangular antiferromagnets; J. Phys. Soc. Jpn.  \textbf{65}, 1799 (1996).

\end{thebibliography}
\end{document}